\UseRawInputEncoding
\documentclass[%
 reprint,
superscriptaddress,
bibnotes,
 amsmath,amssymb,
 aps, prb, 
]{revtex4-2}

\usepackage{graphicx}
\usepackage{dcolumn}
\usepackage{bm}
\usepackage{braket}
\usepackage[colorlinks=true, allcolors=blue]{hyperref}
\usepackage{cleveref}

\begin{document}
\preprint{APS/123-QED}

\title{Carrier Emission and Capture Competition mediated A(n)BC Recombination Model in Semiconductors with Multi-Level Defects}
\author{Shanshan Wang}
\affiliation{School of Microelectronics, Fudan University, Shanghai 200433, China}
\affiliation{Key Laboratory of Computational Physical Sciences (MOE), Fudan University, Shanghai 200433, China}
\author{Menglin Huang}
\email{menglinhuang@fudan.edu.cn}
\affiliation{School of Microelectronics, Fudan University, Shanghai 200433, China}
\affiliation{Key Laboratory of Computational Physical Sciences (MOE), Fudan University, Shanghai 200433, China}
\author{Su-Huai Wei}
\affiliation{Eastern Institute of Technology, Ningbo 315200, China}
\author{Xin-Gao Gong}
\affiliation{Key Laboratory of Computational Physical Sciences (MOE), Fudan University, Shanghai 200433, China}
\author{Shiyou Chen}
\email{chensy@fudan.edu.cn}
\affiliation{School of Microelectronics, Fudan University, Shanghai 200433, China}
\affiliation{Key Laboratory of Computational Physical Sciences (MOE), Fudan University, Shanghai 200433, China}
\date{February 23, 2025}

\begin{abstract}
The ABC model has been widely used to describe the carrier recombination rate, in which the rate of non-radiative recombination assisted by deep-level defects is assumed to depend linearly on excess carrier density $\Delta n$, leading to a constant recombination coefficient A. However, for multi-level defects that are prevalent in semiconductors, we demonstrate here that the rate should depend nonlinearly on $\Delta n$. When $\Delta n$ varies, the carrier capture and emission of defects can change the defect density distribution in different charge states, which can further change the carrier capture and emission rates of the defects and thus make the recombination rate depend non-linearly on $\Delta n$, leading to an $A(n)$ function. However, in many recent calculation studies on carrier recombination rate of multi-level defects, only carrier capture was considered while carrier emission from defect levels was neglected, causing incorrect charge-state distribution and misleading linear dependence of the rate on $\Delta n$. For $\text{V}_{\text{Ga}}$-$\text{O}_{\text{N}}$ in GaN and $\text{Pb}_\text{I}$ in CsPbI$_3$, our calculations showed that neglecting the carrier emission can cause the recombination rate underestimation by more than 8 orders of magnitude when $\Delta n$ is $10^{15}$ cm$^{-3}$. Our findings suggest that the recent studies on carrier recombination assisted by multi-level defects should be revisited with carrier emission considered, and the widely-used $ABC$ model should be reformed into the $A(n)BC$ model.
\end{abstract}

\maketitle


\section{\label{sec:level1}Introduction}

\par Carrier recombination in semiconductors is an important process determining the performances of many semiconductor devices, such as the power dissipation of electronic devices \cite{Lutz2018book, Zhang2010IEEE}, quantum efficiency of light-emitting diodes \cite{Dong2015science}, power conversion efficiency of solar cells \cite{park2018NRM}, and responsivity of photodetectors \cite{Binet1996APL}. In different semiconductors, the carrier recombination can be contributed by different mechanisms (or through different pathways), such as defect-assisted recombination, band-to-band recombination and Auger recombination \cite{zhang2020AEM,wang2022NCS}. The recombination kinetics of these three mechanisms are usually described by the well-known ABC model, which claims the overall recombination rate $R=An+Bn^2+Cn^3$ \cite{Johnston2016ACR,Strauss1993APL,shcherbakov2017NC,kiligaridis2021NC,ball2016NE,zhang2022LSA,chuang2012book,Shen2007APL,li2024Nature},  where $n$ is the carrier density, $A$, $B$ and $C$ represent the monomolecular recombination coefficient (which is typically related to defect-assisted recombination \cite{manser2014NP}), bimolecular recombination coefficient (band-to-band recombination \cite{zhang2020AEM}), and trimolecular recombination coefficient (Auger recombination \cite{Johnston2016ACR}), respectively.  

\par In the $ABC$ model, defect-assisted recombination rate is assumed to have a linear dependence on carrier density. Therefore, the $A$ coefficient is considered as a constant independent of carrier density, which is widely accepted in the carrier dynamics characterization experiments, such as transient spectroscopy and time-resolved photoluminescence measurements \cite{Milot2015AFM,Chen2018AM,saba2014NC}. In these measurements, a constant $A$ is extracted from the exponential fitting of the decay dynamics, for example, $A$ was measured to be constants of $10^7$ s$^{-1}$ in InGaN \cite{Shen2007APL}, $1.8\times10^7$ s$^{-1}$ in CH$_3$NH$_3$PbI$_3$ \cite{Yamada2014JACS}, and $4\times10^7$ s$^{-1}$ in GaAs \cite{Yablonovitch1987APL}. Such a constant $A$ and the $ABC$ model are important for evaluating the contribution of different recombination mechanisms as well as manipulating carrier dynamics and lifetime in semiconductor devices. For example, the $ABC$ model shows that defect-assisted nonradiative recombination (the $An$ term) dominates at low carrier densities, band-to-band radiative recombination (the $Bn^2$ term) dominates at medium carrier densities, and Auger nonradiative recombination (the $Cn^3$ term) governs at high carrier densities, so the design of high-efficiency light-emitting diodes (LED) generally requires that the carrier density should be controlled at a medium level to promote the the radiative $Bn^2$ term and thus achieve the highest quantum efficiency $Bn$/$(A+Bn+Cn^2)$ \cite{liu2021NM,manser2014NP,chen2023NP,David2010APL}. Similar to the case in LED devices, the design of other devices also adopts the ABC model as a basic rule for the carrier dynamics manipulation \cite{Iwata2015JAP,Johnston2016ACR,Herz2016ARPC,deQuilettes2019CR,Kioupakis2013NJP}.

\begin{figure}[htbp]
\centering
\includegraphics[width=0.48\textwidth]{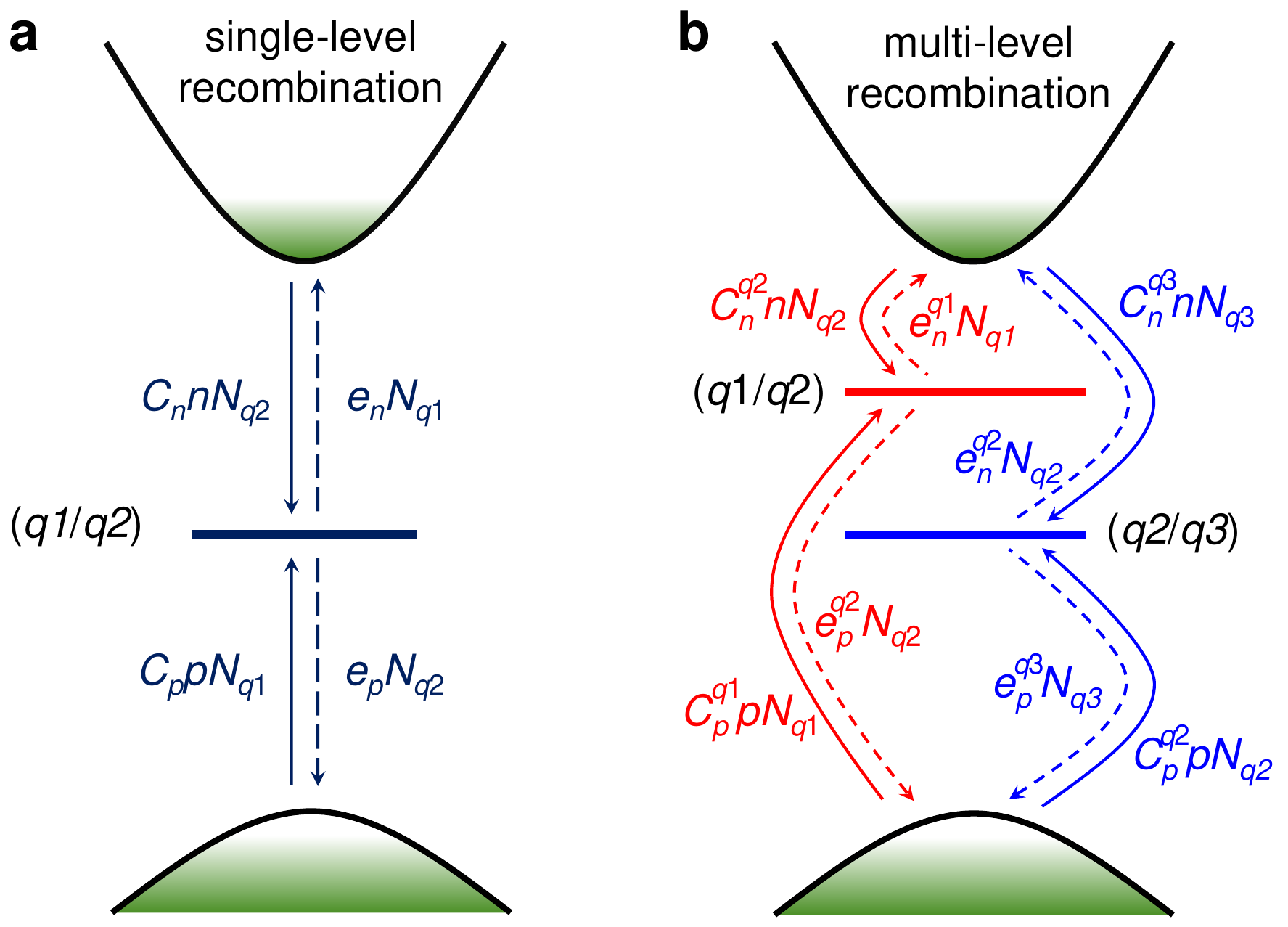}
\caption{Illustration of carrier capture and emission processes during defect-assisted recombination for (a) single-level defect and (b) two-level defect. The solid lines represent the carrier capture process, and the dashed lines represent the carrier emission process. The expressions of the corresponding capture and emission rates are indicated.}\label{Fig1}
\end{figure}

\par Actually, the linear dependence of defect-assisted recombination rate on carrier densities in the $ABC$ model can be traced back to the Shockley-Read-Hall (SRH) model \cite{Shockley1952PR,Hall1952PR}, as shown in Fig. \ref{Fig1}(a). For a defect with a charge-state transition level $(q1/q2)$ in the band gap, there are four processes contributing to the overall carrier recombination: electron capture to the defect level, electron emission from the level, hole capture to and hole emission from the level. For instance, electron capture from CBM to the $(q1/q2)$ defect level can make the defect transit from $q2$ to $q1$ charge state. The corresponding electron capture rate can be expressed as $C_nnN_{q2}$, in which $C_n$ is the capture coefficient, $n$ is the electron carrier density, and $N_{q2}$ is the defect density in $q2$ state. Under continuous external injections or excitation of carriers, the four processes will reach a steady state when the net electron capture rate (capture rate offset by emission rate) equals to the net hole capture rate \cite{Bridgman1928PR}. Consequently, as derived by Shockley, Read and Hall and introduced in many textbooks \cite{Shockley1952PR, Hall1952PR, Landsberg1992book,stoneham2001book}, the defect-assisted recombination rate at steady-state can be expressed as,
\begin{equation}\label{SRH_original}
    R_{\text{SRH}}=\frac{N_{\text{tot}}C_nC_p(np-n_i^2)}{C_n(n+n_1)+C_p(p+p_1)},
\end{equation}
where $N_{\text{tot}}$ is the defect density, $C_n$ and $C_p$ are the electron and hole capture coefficients, respectively, $n$ and $p$ are the electron and hole carrier densities, $n_i$ is the intrinsic carrier density, $n_1$ and $p_1$ are the electron and hole densities when the Fermi level coincides with the defect level. According to Eq. (\ref{SRH_original}), when the defect level is shallow, either $n_1$ or $p_1$ is extremely large, leading to a low value of $R_{\text{SRH}}$. Only when the defect level is deep, can the defect have large $R_{\text{SRH}}$. Therefore, it is well-known that deep-level defects are effective recombination centers \cite{Das2020PRM,Alkauskas2016JAP,Wickramaratne2016APL}. For deep-level defects, $n_1$ and $p_1$ are very small and can be neglected, and usually $n\approx p$ in optoelectronic devices with high excess carrier density ($n\approx p=\Delta n \gg n_0,p_0$, where $n_0$, $p_0$ are equilibrium carrier densities, and $\Delta n$ is the excess carrier density), then the recombination rate can be simplified as $R_{\text{SRH}}=N_{\text{tot}}(C_nC_p)/(C_n+C_p)n$ \cite{Alkauskas2016PRB,Dreyer2016APL,Yuan2022AM}. Compared with the defect-assisted recombination ($An$) term in the $ABC$ model, we can find $A=N_{\text{tot}}(C_nC_p)/(C_n+C_p)$, which is indeed a constant. This is the reason why the linear dependence of defect-assisted recombination rate on carrier density as well as the $ABC$ model work well in dealing with the carrier recombination in deep-level defect systems \cite{Dreyer2016APL,Armstrong2015JAP}.

\par However, it should be noted that the derivation of SRH model was based on a single-level defect \cite{Sah1967IEEE,Shockley1952PR,Hall1952PR}, so the SRH model should be in principle applicable only for defects with one single level in the band gap. Defects in real semiconductors usually produce two or more charge-state transition levels in the band gap, for instance, Si vacancy ($\text{V}_\text{Si}$) defect produces three levels in the 1.17 eV band gap of Si (at 0.38 eV for $0/+2$, 0.63 eV for $0/-1$, and 1.0 eV for $-1/-2$ charge state transition) \cite{Weber2013PRB}, $\text{C}_\text{N}$ in GaN \cite{Lyons2014PRB} and $\text{I}_\text{i}$ in CH$_3$NH$_3$PbI$_3$ \cite{Du2015JPCL} each produces two levels. Each charge-state transition level of such defects can cause carrier capture and emission, determining the overall defect-assisted recombination rate collectively. For multi-level defects, whether the recombination rate still has a linear dependence on carrier densities remains unclear.

\par Given the prevalence of multi-level defects in semiconductors, accurately calculating the recombination rate in multi-level defect systems has been a long-standing research topic over the past 60 years \cite{Sah1958PR,Choo1970PRB,Tyan1992Jxx}. In 2016, Alkauskas et al. studied the recombination in a three-level defect system and defined a total capture coefficient $C_{\text{tot}}$ to calculate its recombination rate \cite{Alkauskas2016PRB}. In the following years, $C_{\text{tot}}$ has been extensively adopted to calculate the recombination rate in multi-level-defect systems \cite{Zhang2022ACIE,Liang2022JACS,Ji2023ACIE,Dou2023PRA,Zhang2023PRXE,Kumgai2023PRXE,Zhang2023JRCL,Liang2022SolarRRL,Tang2024PRM}. For a two-level defect, which has three charge states $-1$, $0$ and $+1$, $C_{\text{tot}}$ was derived as \cite{Zhang2022ACIE,Liang2022JACS,Ji2023ACIE,Dou2023PRA,Zhang2023PRXE,Kumgai2023PRXE,Zhang2023JRCL,Liang2022SolarRRL,Tang2024PRM},
\begin{equation}\label{Ctot}
    C_{\text{tot}}=\frac{C_n^0+C_p^0}{1+\frac{C_n^0}{C_p^-}+\frac{C_p^0}{C_n^+}},
\end{equation}
where $C_n^0$ and $C_p^-$ are the electron and hole capture coefficients of $(0/-1)$ transition level, $C_n^+$ and $C_p^0$ are the electron and hole capture coefficients of $(+1/0)$ transition level. The corresponding recombination rate $R=N_{\text{tot}}C_{\text{tot}}n$. In the derivation of Eq. (\ref{Ctot}), the carrier capture processes at ($+1/0)$ and $(0/-1)$ transition levels were considered, while the corresponding carrier emission processes were neglected. Therefore, the carrier capture rate at each level is directly regarded as the net carrier capture rate, neglecting the contribution of the carrier emission rate. According to Eq. (\ref{Ctot}), $C_{\text{tot}}$ is obviously a constant independent of carrier density. Thus, the $A$ coefficient ($A=N_{\text{tot}} C_{\text{tot}}$) in the $ABC$ model is also a constant, which indicates that the recombination rate for multi-level defect also has a linear dependence on carrier density.

\par However, in this work we will show the $A$ coefficient is actually NOT a constant in semiconductors with multi-level defects. The competition between carrier capture and carrier emission processes influences the distribution of defect density in different charge states and thus the overall recombination rate significantly. Since the competition varies with excess carrier densities, the $A$ coefficient changes quickly as a function of the carrier density $n$, rather than being a constant. In recent calculation studies based on $C_{\text{tot}}$ formula, neglecting the carrier emission from defect levels make the defects act only as carrier trapping centers and stay ineffective for carrier recombination, therefore, the carrier recombination rates can be severely underestimated by orders of magnitude and the coefficient behaves as a constant erroneously. Our results indicate that when multi-level defects exist, a more rigorous formula considering both the carrier capture and emission of all defect levels should be used for calculating the recombination rate $A(n)n$, and the well-known $ABC$ model should be reformed into the $A(n)BC$ model in carrier dynamics manipulation.

\section{\label{sec:level1}Results}

\subsection{\label{sec:level2}Sah-Shockley recombination model of multi-level defects}

\par The carrier recombination assisted by multi-level defect has been rigorously studied by Sah and Shockley in 1958 \cite{Sah1958PR}, where the derivation of recombination rate included both carrier capture and emission. Unfortunately, recent studies followed the rate equation of single deep-level defect and neglected carrier emission contribution, deriving the simplified formula of a total capture coefficient $C_{\text{tot}}$, e.g., Eq. (\ref{Ctot}) \cite{Zhang2022ACIE,Liang2022JACS,Ji2023ACIE,Dou2023PRA,Zhang2023PRXE,Kumgai2023PRXE,Zhang2023JRCL,Liang2022SolarRRL,Tang2024PRM}. Here, to demonstrate the importance of carrier emission, and show the difference of recombination assisted by single-level and multi-level defects, we follow the derivation of Sah and Shockley and introduce the recombination rate formula first.

\par In multi-level defect systems, the carrier capture and emission processes may occur individually at different levels, complicating the overall recombination process. For simplicity, we illustrate in Fig. \ref{Fig1}(b) by a two-level defect with three charge states $q1$, $q2$, $q3$, where the $(q1/q2)$ charge-state transition level is close to the conduction band minimum (CBM) while the $(q2/q3)$ level is located deeply in the band gap. There are totally 8 capture and emission processes at two levels, which collaboratively determine the overall recombination rate. The capture and emission rates have been indicated in Fig. \ref{Fig1}(b). For carrier emission, the rate constant $e_n$ ($e_p$) can be expressed as $C_nn_1$ ($C_pp_1$) or $C_nn_2$ ($C_pp_2$), where $n_1$ and $p_1$ are the electron and hole densities when the Fermi level coincides with the defect $(q1/q2)$ level, and $n_2$ and $p_2$ are those when the Fermi level coincides with the defect $(q2/q3)$ level. Under steady-state condition, the net electron capture rate should be equal to the net hole capture rate for each level,
\begin{equation}\label{sah-shockley1}
    C_n^{q2} nN_{q2}-C_n^{q2} n_1 N_{q1}=C_p^{q1} pN_{q1}-C_p^{q1} p_1 N_{q2},  
\end{equation}
\begin{equation}\label{sah-shockley2}
    C_n^{q3} nN_{q3}-C_n^{q3} n_2 N_{q2}=C_p^{q2} pN_{q2}-C_p^{q2} p_2 N_{q3}, 
\end{equation}
where $C_n^{q2}$ and $C_p^{q1}$ are the electron and hole capture coefficients of $(q1/q2)$ level, and $C_n^{q3}$ and $C_p^{q2}$ are those of $(q2/q3)$ level. Combining the condition that the total defect density should be equal to the defect density in all the three charge states, $N_{\text{tot}} = N_{q1} + N_{q2} + N_{q3}$, with Eqs. (\ref{sah-shockley1}, \ref{sah-shockley2}), the defect density in each charge state under the steady-state condition can be expressed as,
\begin{equation}\label{Nq2}
    N_{q2}=DN_{\text{tot}},
\end{equation}

\begin{equation}\label{Nq3}
    N_{q3}=D \frac{C_p^{q2}p+C_n^{q3}n_2}{C_n^{q3} n+C_p^{q2} p_2} N_{\text{tot}},
\end{equation}
\begin{equation}\label{Nq1}
    N_{q1}=D \frac{C_n^{q2}n+C_p^{q1}p_1}{C_p^{q1} p+C_n^{q2} n_1} N_{\text{tot}},
\end{equation}
where $D=(1+\frac{C_p^{q2}p+C_n^{q3}n_2}{C_n^{q3} n+C_p^{q2} p_2}+\frac{C_n^{q2}n+C_p^{q1}p_1}{C_p^{q1} p+C_n^{q2} n_1})^{-1}$. Consequently, the recombination rate can be written as \cite{Sah1958PR} (the detailed derivations and the extension to arbitrary number of defect levels are given in Supplemental Material \cite{SM,Zhang2020JPCC,Moustafa2024PRM}),
\begin{widetext}
\begin{equation}\label{R_EIR}
    R_{\text{EIR}}=N_{\text{tot}} \frac{\left( C_n^{q2}+C_n^{q3}\frac{C_p^{q2}p+C_n^{q3}n_2}{C_n^{q3}n+C_p^{q2}p_2} \right)n-C_n^{q2}\frac{C_n^{q2}n+C_p^{q1}p_1}{C_p^{q1}p+C_n^{q2}n_1}n_1-C_n^{q3}n_2}{1+\frac{C_p^{q2}p+C_n^{q3}n_2}{C_n^{q3}n+C_p^{q2}p_2}+\frac{C_n^{q2}n+C_p^{q1}p_1}{C_p^{q1}p+C_n^{q2}n_1}}.
\end{equation}    
\end{widetext}
\par Following the definition of $A$ \cite{wang2022NCS}, the multi-level defect-assisted recombination coefficient can be defined as,
\begin{equation}\label{A_EIR}
    A_{\text{EIR}} = \frac{R_{\text{EIR}}}{\Delta n}.
\end{equation}
Since the carrier emission has been included in the derivation, we denote it as emission-included recombination (EIR). If the carrier emission is completely neglected, the rate of emission-omitted recombination (EOR) can be simplified from Eq. (\ref{R_EIR}) as,
\begin{equation}\label{R_EOR}
    R_{\text{EOR}}=N_{\text{tot}} \frac{C_n^{q2} n+C_p^{q2} p}{1+\frac{C_p^{q2} p}{C_n^{q3} n}+\frac{C_n^{q2} n}{C_p^{q1} p}}.
\end{equation}
The $A_{\text{EOR}}$ coefficient can also be defined as $A_{\text{EOR}} = R_{\text{EOR}}/\Delta n$, similar to Eq. (\ref{A_EIR}). Note that if a condition of $n = p$ is supposed, Eq. (\ref{R_EOR}) can be further simplified to “$C_{\text{tot}}$” formula proposed by Alkauskas et al., that is, $R=N_{\text{tot}} C_{\text{tot}} n$, where $C_{\text{tot}}$ is given in Eq. (\ref{Ctot}). Below we will take an example to show the difference between $A_{\text{EIR}}$ and $A_{\text{EOR}}$, and illustrate that $A$ is actually not a constant in nonradiative recombination assisted by multi-level defects.

\begin{figure*}[htbp]
\centering
\includegraphics[width=0.8\textwidth]{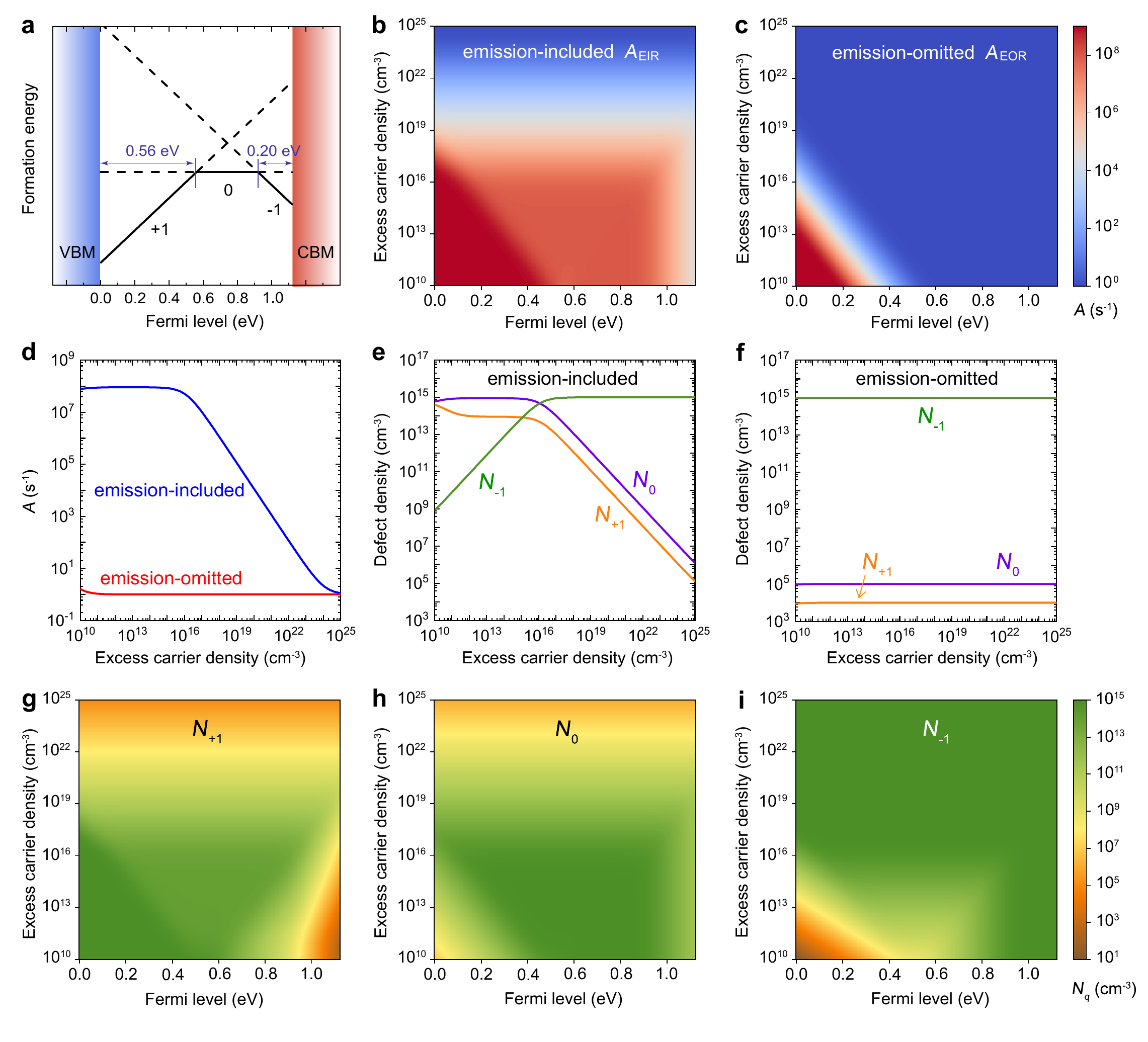}
\caption{\textbf{Illustration of the effect of carrier emission on the defect-assisted recombination in a two-level defect system.} (a) Formation energy of a defect with two levels in the band gap, (+1/0) and (0/-1), as a function of Fermi level. (b) Calculated A coefficient as a function of equilibrium Fermi level and excess carrier density, when carrier emission is included. (c) Calculated A coefficient as a function of equilibrium Fermi level and excess carrier density, when carrier emission is omitted. (d) Comparison of the emission-included and emission-omitted A coefficients as functions of excess carrier density. Here the equilibrium Fermi level is assumed to be located at the middle of the band gap. (e) Calculated defect densities in 0, -1 and +1 charge states as functions of excess carrier density, when emission is included. (f) Calculated defect densities in different charge states as functions of excess carrier density, when emission is omitted. (g-i) Calculated defect densities in 0, -1 and +1 charge state as functions of equilibrium Fermi level and excess carrier density, when emission is included.}\label{Fig2}
\end{figure*}

\subsection{\label{sec:level2}Carrier emission contribution to recombination rate}
\par We use a two-level defect in a semiconductor with 1.12 eV band gap as an example. The defect introduces two charge-state transition levels in the band gap, $(+1/0)$ and $(0/-1)$, as shown in Fig. \ref{Fig2}(a). The relatively shallower $(0/-1)$ level is 0.2 eV below the CBM, and the deep $(+1/0)$ level is located at 0.56 eV above the VBM, corresponding to the $(q1/q2)$ and $(q2/q3)$ transition levels shown in Fig. \ref{Fig1}(b). Such defects are common in semiconductors and similar to impurities in silicon, such as Au and Pt \cite{Kwon1987JAP,Watkins1991PRL}. The capture coefficients $C_n^+$ and $C_p^0$ at $(0/+)$ level are $10^{-6}$ cm$^3$ s$^{-1}$ and $10^{-7}$ cm$^3$ s$^{-1}$, $C_n^0$ and $C_p^-$ at $(0/-1)$ level are $10^{-5}$ cm$^3$ s$^{-1}$ and $10^{-15}$ cm$^3$ s$^{-1}$, respectively. These coefficients follow the general trend in the nonradiative multiphonon theory that the nonradiative capture rate increases exponentially as the electronic transition energy between the initial and final states decreases \cite{Alkauskas2014PRB,Zhao2023PRL,abakumov1991book,Zhou2025arxiv}. The total defect density is assumed to be $10^{15}$ cm$^{-3}$. The defect densities distributed in the three charge states as well as the overall recombination rates can be calculated following Eqs. (\ref{sah-shockley1}-\ref{R_EOR}) where $q1$, $q2$, and $q3$ are referred as $-1$, $0$, and $+1$, respectively.

\par Fig. \ref{Fig2}(b, c) show the heat maps of calculated $A$ coefficients as functions of excess carrier density $\Delta n$ and equilibrium Fermi level $E_F$ when the carrier emission contribution is included and omitted, respectively. As $E_F$ shifts from the left side (VBM) to right side (CBM), the electrical conductivity changes from $p$-type to intrinsic, and then to $n$-type. From the heat map, we can see that including the carrier emission contribution has very large influence on the $A$ coefficients, i.e., the emission-included $A_{\text{EIR}}$ is larger than the emission-omitted $A_{\text{EOR}}$ in most areas of the map. Below we will introduce the differences in detail and analyze their physical origins.

\par When the equilibrium Fermi level is located at the middle of the band gap (corresponding to $n_0=p_0$), we plot in Fig. \ref{Fig2}(d) the variation of $A$ coefficients with excess carrier density. According to $A_{\text{EOR}} = R_{\text{EOR}}/\Delta n$, Eq. (\ref{R_EOR}) where $n=p=n_0+\Delta n \approx ∆n$, and Eq. (\ref{Ctot}), we can derive that $A_{\text{EOR}} \approx N_{\text{tot}} C_{\text{tot}}$, so $A_{\text{EOR}}$ is almost independent of excess carrier densities, explaining the nearly constant straight line of $A_{\text{EOR}}$ in Fig. \ref{Fig2}(d). Considering the values of $C_n^+$, $C_p^0$, $C_n^0$ and $C_p^-$, it is easy to understand that the calculated $A_{\text{EOR}}$ is very low (on the order of $10^0$ s$^{-1}$). Therefore, such a defect is not an effective recombination center when the carrier emission is omitted. In contrast, $A_{\text{EIR}}$ exhibits a strong dependence on excess carrier density, i.e., it changes by 7-8 orders of magnitude as the excess carrier density increases in Fig. \ref{Fig2}(d).  When carrier density is in the range 10$^{10}$-10$^{16}$ cm$^{-3}$, $A_{\text{EIR}}$ can be as large as 10$^8$ s$^{-1}$, indicating the defect can be a very effective recombination center, which contradicts the conclusion obtained from $A_{\text{EOR}}$ when carrier emission is omitted. Only when the excess carrier density is extremely high ($>10^{24}$ cm$^{-3}$), can $A_{\text{EIR}}$ be close to $A_{\text{EOR}}$. These results indicate that the $A_{\text{EIR}}$ and $A_{\text{EOR}}$ coefficients may differ by 7-8 orders of magnitude for solar cell (usually $\Delta n\approx$ 10$^{15}$ cm$^{-3}$) \cite{zhang2020AEM} and 5-6 orders of magnitude for light-emitting diode (usually $\Delta n\approx$ 10$^{18}$ cm$^{-3}$) applications \cite{David2010APL}.

\par To reveal the origin of the large discrepancy between $A_{\text{EIR}}$ and $A_{\text{EOR}}$, we show in Fig. \ref{Fig2}(e, f) the defect densities distributed in $+1$, $0$ (neutral), $-1$ charge states. When the emission effect is included in the calculation, the densities in the $0$ and $+1$ charge states keep almost unchanged at a high value around 10$^{15}$ cm$^{-3}$ (almost all the defects are in these two charges states) when $\Delta n$ is low, but decrease rapidly to a very low value when $\Delta n$ is high, as shown in Fig. \ref{Fig2}(e). Such a trend is quite similar to that of the emission-included $A_{\text{EIR}}$ in Fig. \ref{Fig2}(d), indicating that the recombination coefficient is positively and strongly correlated with the defects in the $0$ and $+1$ charge states. The defect density in the $-1$ charge state show an opposite trend to those of the $0$ and $+1$ charge states. When the carrier emission is omitted, the density distribution become completely different, and almost all the defects take the $-1$ charge state, as shown in Fig. \ref{Fig2}(f). The densities of all the three charge states are constants independent of the carrier density, in accordance with the constant $A_{\text{EOR}}$ in Fig. \ref{Fig2}(d), further indicating that the recombination coefficient of the two-level defect is strongly correlated with its density distribution in different charge states.

\par Now we will analyze how the charge-state distribution of defects are correlated with the carrier emission and recombination coefficient. As shown in Fig. \ref{Fig2}(a), there are two charge state transition levels, $(0/-1)$ and $(0/+1)$, in the band gap. The $(0/-1)$ transition level is high and close to CBM. Therefore, the transition energy of the electron capture from CBM to $(0/-1)$ level is small and thus the electron capture coefficient $C_n^0$ is large, while the transition energy of the hole capture from VBM to $(0/-1)$ level is large and thus the hole capture coefficient $C_p^-$ is very small \cite{Henry1977PRB}. Correspondingly, $n_1$ is large and $p_1$ is small.  According to Eq. (\ref{SRH_original}), the single $(0/-1)$ level has a very small recombination coefficient and thus cannot assist the electron-hole recombination effectively. In contrast, the $(0/+1)$ transition level is deep in the middle of the band gap, so its electron and hole capture coefficients, $C_n^+$ and $C_p^0$, both have medium values, and both $n_2$ and $p_2$ are not very large. Therefore, the single deep $(0/+1)$ level can have a large recombination coefficient, in accordance with the common expectation that the deep-level defects are effective recombination centers in the SRH theory. Since only the deep $(0/+1)$ level can assist electron-hole recombination effectively while the $(0/-1)$ level cannot, we can understand why the strong and positive correlation between the recombination coefficient $A$ in Fig. \ref{Fig2}(d) and the $0$ and $+1$ charged defect densities in Fig. \ref{Fig2}(e), because the $0$ and $+1$ charged defects can be involved directly in electron-hole recombination. The $-1$ charge defects cannot be involved in the recombination directly, and can only be involved after it emits an electron or captures a hole and then transits back into the neutral state. Therefore, the coefficient $A$ in Fig. \ref{Fig2}(d) is inversely correlated with the $-1$ charged defect density in Fig. \ref{Fig2}(e).

\par As we can see, the neutral state can participate in both the $(0/-1)$ and $(0/+1)$ transitions, so it is special in the three states and can mediate the carrier capture and emission rates of the two levels, making the two levels influence each other during the electron-hole recombination. The influence gives rise to the obvious changes of the defect charge state distribution in Fig. \ref{Fig2}(e) and the changes of recombination coefficient $A$ in Fig. \ref{Fig2}(d).  As the excess carrier excitation or injection reaches a steady state, the electron and hole capture/emission at the shallower $(0/-1)$ level satisfy Eq. (\ref{sah-shockley1}), thus
\begin{equation}\label{summed_equation}
    (C_n^0 n+C_p^- p_1)N_0=(C_p^- p+C_n^0 n_1)N_{-1},
\end{equation}
which means that the summed transition rate (electron capture plus hole emission) from neutral to $-1$ charged state equals that (hole capture plus electron emission) from $-1$ charged to neutral state. Since $C_p^-p_1$ is much smaller than $C_n^0 n$, and $C_p^- p$ is smaller than $C_n^0 n_1$, the hole capture and hole emission rates can be reasonably neglected, giving $C_n^0 nN_0 \approx C_n^0 n_1 N_{-1}$. Hence, the electron capture rate $C_n^0 n$ and its counteracting electron emission rate $C_n^0 n_1$ are dominant in Eq. (\ref{summed_equation}). When the excess carrier density is low (thus $n$ is low), $C_n^0 n$ is much smaller than $C_n^0 n_1$, then Eq. (\ref{summed_equation}) requires the neutral defect density $N_0$ to be much higher than the $-1$ charged defect density $N_{-1}$. The high density of neutral defects leads to high recombination coefficient through the $(0/+1)$ deep level. When the excess carrier density increases to a high level (thus $n$ is high), $C_n^0 n$ becomes comparable to or even higher than $C_n^0 n_1$, then $N_0$ has to decrease to be comparable to or lower than $N_{-1}$. The decrease of $N_0$  further causes the decrease of the recombination coefficient $A$. Based on such a competition between the electron capture and emission rate associated with the $(0/-1)$ level, we can explain the decrease of $N_0$ in Fig. \ref{Fig2}(e) and $A$ in Fig. \ref{Fig2}(d).

\par If the carrier emission effect is omitted in the calculation, as shown in Eq. (\ref{R_EOR}) and reported in the calculations based on the $C_{\text{tot}}$ formula, the competition disappears completely, then Eq. (\ref{summed_equation}) cannot be satisfied. Since the capture rate $C_n^0n$ of $(0/-1)$ level is high, the defects capture electrons quickly, transits into the $-1$ charge states and cannot transit back to the neutral state any more. That makes almost all the defects take the $-1$ charge state, giving a high and constant $N_{-1}$ and very low $N_0$ and $N_{+1}$ independent of excess carrier density in Fig. \ref{Fig2}(f). The constant and low density of neutral and $+1$ charged defects makes the recombination coefficient $A_{\text{EOR}}$ also lose its dependence on excess carrier densities and much lower than $A_{\text{EOR}}$ in Fig. \ref{Fig2}(d).

\par Taking the intrinsic system with a middle-gap Fermi level ($n_0=p_0$) as an example, the analysis above shows in detail how omitting the carrier emission effect in the calculation causes the incorrect defect charge-state distribution and thus the seriously underestimated recombination coefficient $A$. Besides, for $n$-type or $p$-type system where the Fermi level approaches CBM or VBM, omitting carrier emission also causes large difference in the calculated coefficients $A_{\text{EIR}}$ and $A_{\text{EOR}}$. The corresponding distribution of defect densities with emission included are shown in Fig. \ref{Fig2}(g-i), while those with emission omitted are shown in Fig. S1 in Supplemental Material. We can also see that the very different $A_{\text{EIR}}$ and $A_{\text{EOR}}$ are correlated with the large differences in defect-density distribution. Detailed analyses are presented in Supplemental Material.

\par According to a series of recent first-principles calculations based on the $C_{\text{tot}}$ formula that omitted the carrier emission, the multi-level defects with both shallow and deep levels are mostly stuck at one charge state associated only to the shallow level. This is because the shallow level typically has a high capture rate for one type of carriers, while has a very low capture rate for the other type of carriers. Since the overall electron-hole recombination rate is limited by the lower rate, the slower carrier capture process is denoted as a “rate-limiting” step in previous studies \cite{Zhang2020JPCC,Dreyer2016APL}. For example, in our example shown above, omitting carrier emission shows most defects are stuck at $-1$ charge state, so the slow hole capture process at $(0/-1)$ level acts as the “rate-limiting” step, which erroneously deactivates this defect in the overall recombination process. However, when carrier emission is included, we find it is exactly the carrier emission from the shallower $(0/-1)$ level that reactivates the defects from $-1$ charge state to the neutral state and then assists the electron-hole recombination through the deep $(0/+1)$ level. Therefore, our results show that the carrier emission from the shallower level can be regarded as an important “rate-reactivating” step, rather than a “rate-limiting” step. 

\par It should be noted that if the recombination at shallow $(0/-)$ level is more efficient than that at deep $(0/+)$ level, probably caused by different phonon frequencies or anharmonicity during multiphonon transition, considering carrier emission would retard the overall recombination. In such a rare scenario, $A_{\text{EIR}}$ can be lower than $A_{\text{EOR}}$ when carrier density is low.

\begin{figure*}[htbp]
\centering
\includegraphics[width=0.85\textwidth]{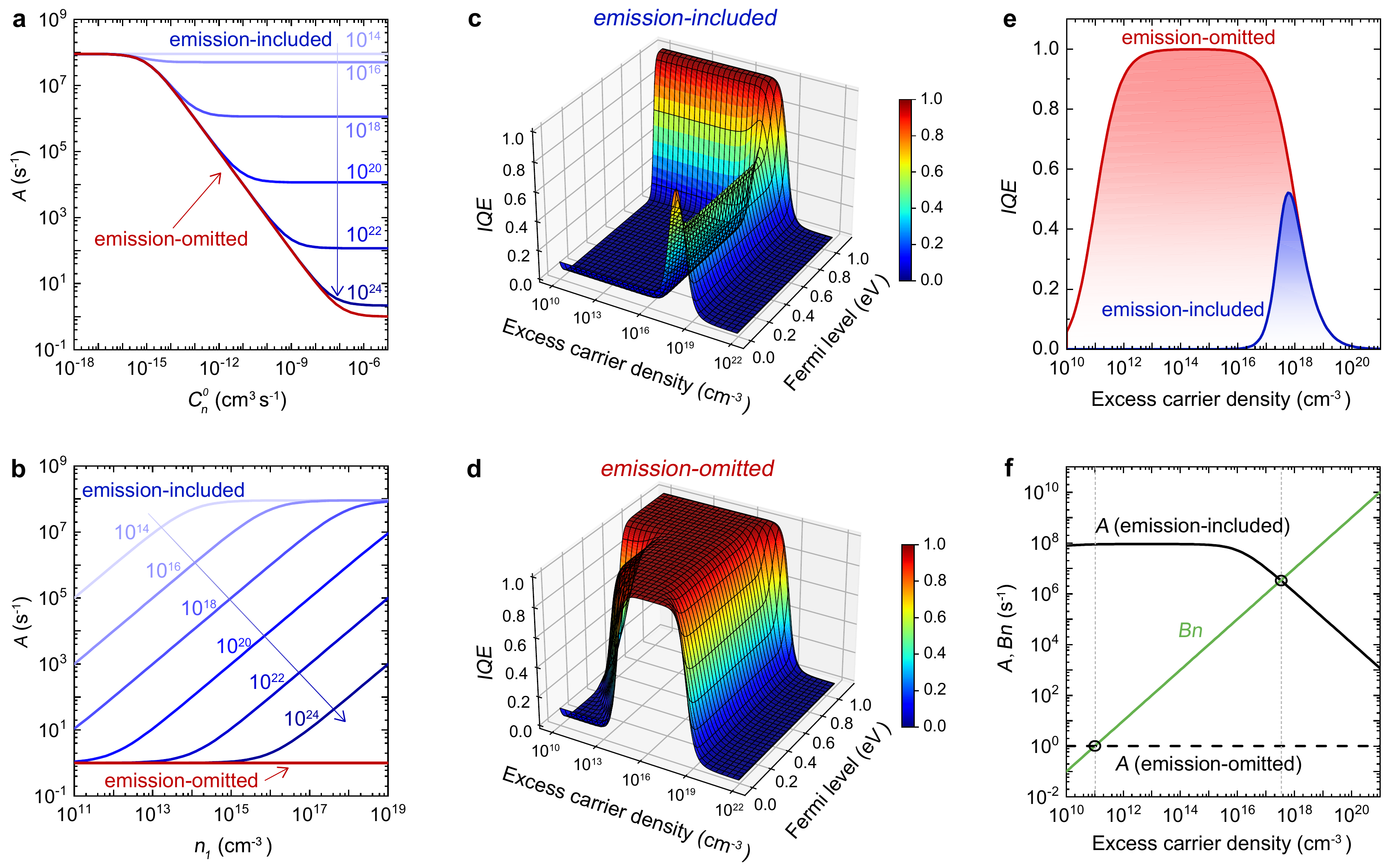}
\caption{\textbf{Comparison of $A_{\text{EIR}}$ and $A_{\text{EOR}}$ coefficients with variable capture coefficients and transition levels, and the influence of carrier emission on internal quantum efficiency} (a) Calculated $A_{\text{EIR}}$ (blue lines) and $A_{\text{EOR}}$ (red line) coefficients as functions of electron capture coefficient $C_n^0$. (b) Calculated $A_{\text{EIR}}$ (blue lines) and $A_{\text{EOR}}$ (red line) coefficients as functions of $n_1$ (which is determined by the energy difference between the defect $(0/-)$ level and CBM level). Note that $A_{\text{EIR}}$ varies with excess carrier densities, while $A_{\text{EOR}}$ does not. (c, d) Simulated internal quantum efficiencies based on $A_{\text{EIR}}$ and $A_{\text{EOR}}$, respectively. The total defect density of all charge states is assumed to be 10$^{15}$ cm$^{-3}$. (e) Comparison of internal quantum efficiencies based on $A_{\text{EOR}}$ (red curve) and $A_{\text{EIR}}$ (blue curve) when the Fermi level is at the middle of the band gap. (f) Comparison of the $A$ and $Bn$ terms in the $ABC$ recombination model with different excess carrier density.}\label{Fig3}
\end{figure*}

\subsection{\label{sec:level2}Strict conditions required for omitting carrier emission}

\par Given the emission rate constant has a form as $C_n^0 n_1$, to figure out the condition when carrier emission can be reasonably omitted, we investigate the variation of $A$ coefficients with electron capture coefficient $C_n^0$ in Fig. \ref{Fig3}(a), and the variation of $A$ coefficients with $n_1$ (determined by the energy difference between the defect $(0/-)$ level and CBM level) in Fig. \ref{Fig3}(b). In Fig. \ref{Fig3}(a), we gradually increase the electron capture coefficient $C_n^0$ and meanwhile fix the $(0/-)$ level at a shallow location, 0.2 eV below CBM. When $C_n^0$ is very small, the carrier emission can be omitted because $A_{\text{EIR}}$ and $A_{\text{EOR}}$ are almost equal. As $C_n^0$ increases, the differences between $A_{\text{EIR}}$ and $A_{\text{EOR}}$ are widened due to the enhanced carrier emission effect of $(0/-)$ level. The blue lines show $A_{\text{EIR}}$ with varied excess carrier densities $\Delta n$. When $\Delta n$ is low, e.g., 10$^{14}$ cm$^{-3}$, $C_n^0 n_1$ is much larger than $C_n^0 n$, so the defects are mostly in the neutral state according to Eq. (\ref{summed_equation}) and can give a high recombination rate through the deep $(0/+1)$ level, and thus, $A_{\text{EIR}}$ is a large constant at 10$^{8}$ s$^{-1}$ with varied $C_n^0$. For a higher $\Delta n$, $A_{\text{EIR}}$ decreases as $C_n^0$ increases. Only when ∆n is as high as 10$^{24}$ cm$^{-3}$, the $A_{\text{EIR}}$ and $A_{\text{EOR}}$ lines overlap in the most range of $C_n^0$, indicating that the carrier emission can be neglected under a such high excess carrier density.

\par In Fig. \ref{Fig3}(b), the electron capture coefficient is fixed but $n_1$ varies from 10$^{11}$ cm$^{-3}$ to 10$^{19}$ cm$^{-3}$, i.e., the $(0/-)$ level shifts upward from a deep middle-gap location to a shallower location near CBM. The calculated $A_{\text{EOR}}$ is a constant independent of $n_1$ and the $(0/-)$ level, in accordance to Eq. (\ref{R_EOR}). However, because of the competition between $C_n^0 n_1$ and $C_n^0 n$, $A_{\text{EIR}}$ has a strong dependence on both the location of the $(0/-1)$ level in the band gap and the carrier density. The difference between $A_{\text{EIR}}$ and $A_{\text{EOR}}$ is widened when the $(0/-)$ level becomes shallower ($n_1$ increases) and the excess carrier density decreases. Only when $\Delta n$ is higher than $n_1$ by 9-10 orders of magnitude, can $A_{\text{EIR}}$ and $A_{\text{EOR}}$ lines overlap. That means, either $n_1$ has to been very low or $\Delta n$ has to be very high. Since $n_1$ of a shallow level is usually higher than 10$^{15}$ cm$^{-3}$, $\Delta n$ has to be higher than 10$^{24}$ cm$^{-3}$.

\par Based on these analyses, in the calculation of the recombination coefficient of multi-level defects, omitting the carrier emission can be acceptable only if one of the three conditions are satisfied: (i) the excess carrier density $\Delta n$ is very high, almost 10$^{24}$ cm$^{-3}$. It should be noted that such a high carrier density is beyond the common carrier density in optoelectronic devices, so this strict condition is meaningless for studies of functional semiconductors; (ii) the shallow level of the defect has a very small $C_n^0$ (for shallow levels near CBM) or $C_p^0$ (for shallow levels near VBM). This condition is also not easy to be satisfied, because the small energy difference between the shallow level and the band edge usually leads to a large carrier capture coefficient according to the multiphonon theory; (iii) All levels of the defect are deep (far from the band edge), so even the shallowest level has a very low $n_1$ or $p_1$, e.g., lower than 10$^{10}$ cm$^{-3}$. Since the first and second conditions are both too strict to be satisfied and only the third condition is possible, omitting the carrier emission can cause large errors in the calculated recombination coefficient if one of multiple defect levels is shallow.

\subsection{\label{sec:level2}A(n)BC recombination model and quantum efficiency calculation}

\par The section above shows clearly that the multi-level defect assisted recombination coefficient $A$ is a function of carrier density, $A(n)$, when both the carrier capture and emission are considered. Therefore, the defect-assisted recombination coefficient $A$ in the $ABC$ recombination model $R=An+Bn^2+Cn^3$ should also be replaced by the $A(n)$ function, then the model becomes $R=A(n)n+Bn^2+Cn^3$. In this sense, it becomes questionable to attribute simply the first-order term to the defect-assisted recombination, the second-order term to the band-to-band recombination and the third-order term to the Auger recombination, because the multi-level defect-assisted recombination rate $A(n)n$ is not exactly first-order. Neglecting the carrier emission will cause a constant $A$ coefficient ($A_{\text{EOR}}$) and first-order dependence of the rate on carrier density, which will cause large errors in the calculated total recombination rate. The errors can further cause large errors in the calculation of the internal quantum efficiency (IQE), which is the ratio of radiative recombination rate $Bn^2$ in total recombination rate $R$ \cite{Johnston2016ACR},
\begin{equation}\label{IQE}
    IQE=\frac{Bn}{A+Bn+Cn^2},
\end{equation}
where $A$ is the first term of the denominator. IQE is an important quantity for assessing the optoelectronic performance of light-emitting and photovoltaic semiconductors, so the calculation errors can cause serious problems in the design of LED and solar cell devices.

\par To show how large errors the omission of carrier emission and the constant $A_{\text{EOR}}$ coefficient can cause in the calculation of IQE, we calculate in Fig. \ref{Fig3}(c, d) the variation of IQE with excess carrier density and equilibrium Fermi level. IQE in Fig. \ref{Fig3}(c) is calculated based on the carrier emission included $A_{\text{EIR}}$, and that in Fig. \ref{Fig3}(d) is calculated based on the carrier emission omitted $A_{\text{EOR}}$. The radiative recombination coefficient ($B$ coefficient) is assumed to be 10$^{-11}$ cm$^{3}$ s$^{-1}$, and Auger recombination coefficient ($C$ coefficient) is assumed to be 10$^{-29}$ cm$^{6}$ s$^{-1}$ (typical values for semiconductors) \cite{wang2022NCS}. Comparing Fig. \ref{Fig3}(c) and Fig. \ref{Fig3}(d), it is obvious that neglecting the carrier emission underestimates the $A$ coefficient and thus overestimates the IQE significantly in most areas. 

\par This can be seen more clearly in Fig. \ref{Fig3}(e) which shows the IQE with the Fermi level fixed at the middle of the band gap. The emission-omitted IQE (red curve) is much higher than the emission-included IQE (blue curve). In the region with the excess carrier density 10$^{12}$-10$^{16}$ cm$^{-3}$, the red curve shows a wide peak with almost 100\% efficiency, while the blue curve shows almost no efficiency. Only when the excess carrier density is around 10$^{18}$ cm$^{-3}$, the blue curve shows an efficiency peak of only 50\%. The large difference of IQE results from the difference between $A_{\text{EIR}}$ and $A_{\text{EOR}}$, which changes the competition between the non-radiative recombination term $A$ and the radiative recombination term $Bn$ in Eq. (\ref{IQE}). As shown in Fig. \ref{Fig3}(f), $A_{\text{EOR}}$ is a small constant and becomes much smaller than $Bn$ when excess carrier density is higher than 10$^{12}$ cm$^{-3}$, leading to the very high IQE in Fig. \ref{Fig3}(e). In contrast, the non-radiative $A_{\text{EIR}}$ term is much larger than the $Bn$ term, suppressing IQE to almost 0 when the excess carrier density is lower than 10$^{16}$ cm$^{-3}$. As the density further increases, $A_{\text{EIR}}$ decreases and becomes smaller than the $Bn$ term when the density is higher than $3\times 10^{17}$ cm$^{-3}$, then IQE increases to a higher level. 

The comparison shows clearly that the small and constant $A$ coefficient caused by neglecting carrier emission effect can cause large errors in the IQE calculation based on the $ABC$ recombination model. Not only the peak efficiency, but also the curve shape of the IQE dependence on the excess carrier density and Fermi level, can have large errors if the carrier emission effect is not considered. Such kind of errors and the dependence of $A(n)$ coefficient on excess carrier density in the $A(n)BC$ model should be paid special attention to in the computational design of various optoelectronic devices based on semiconductors with multi-level defects.

\subsection{\label{sec:level2}Revisiting recent calculations based on $\textbf{C}_{\textbf{tot}}$ formula}

\par The discussion above demonstrated that carrier emission plays a critical role in determining the distribution of defect densities in different charge states and thus the overall recombination rate induced by multi-level defects, however, the role is neglected in the recent calculations based on the total capture coefficient $C_{\text{tot}}$ formula which was firstly proposed in 2016 \cite{Alkauskas2016PRB}. We noticed that dozens of recent studies have adopted this formula to analyze the recombination statistics of multi-level defects in materials, and obtained $A$ coefficients that are independent of carrier densities \cite{Zhang2022ACIE,Liang2022JACS,Ji2023ACIE,Dou2023PRA,Zhang2023PRXE,Kumgai2023PRXE,Zhang2023JRCL,Liang2022SolarRRL,Tang2024PRM}. Here we start from the original data in Ref. \cite{Alkauskas2016PRB} and recalculate the recombination coefficients to show that the direct use of the $C_{\text{tot}}$ formula with no carrier emission effect may cause serious underestimations of $A$ coefficients and the defect-assisted non-radiative combination rate.

\begin{figure}[htbp]
\centering
\includegraphics[width=0.5\textwidth]{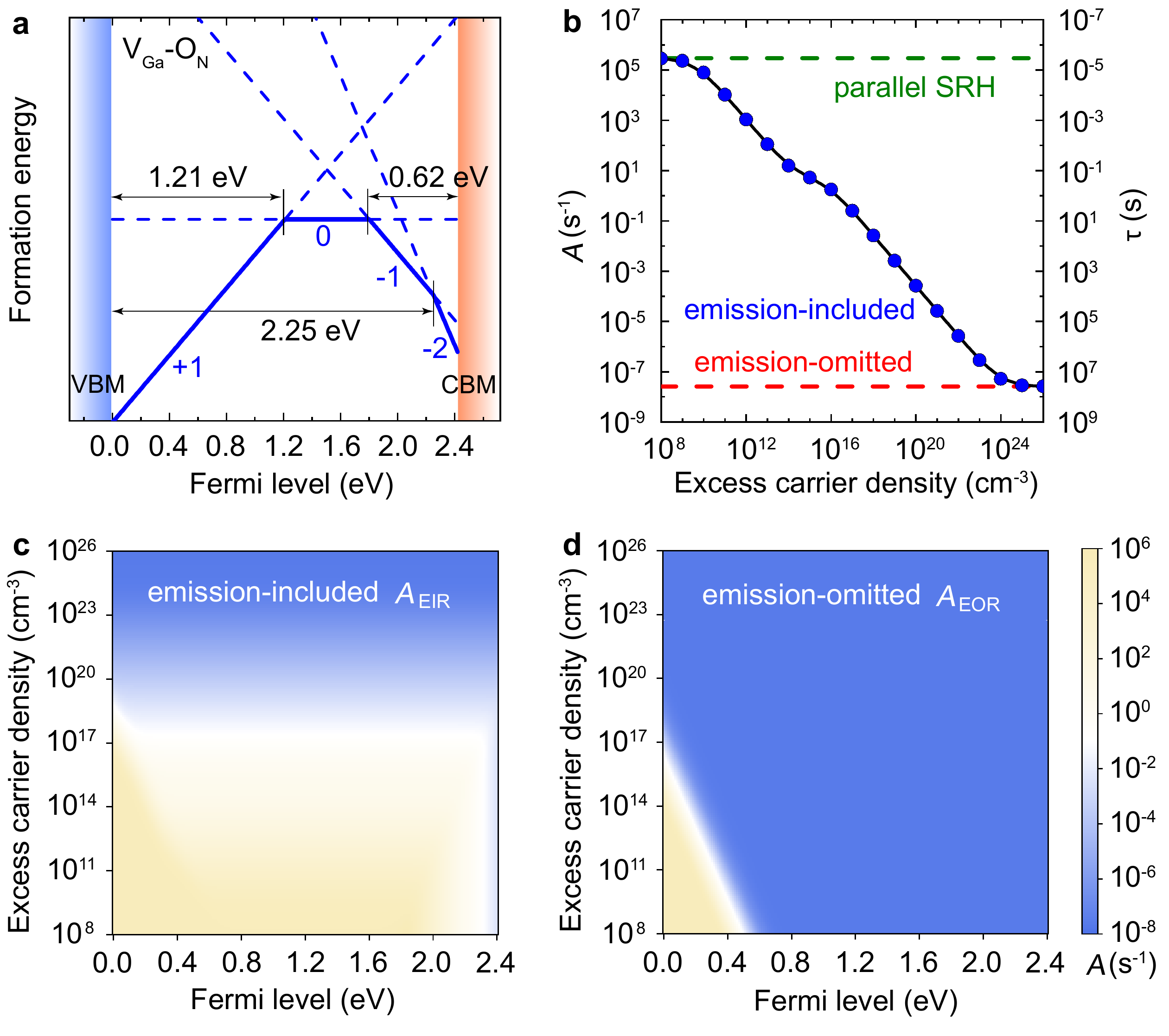}
\caption{\textbf{Effects of carrier emission on the nonradiative recombination assisted by a three-level defect, $\text{V}_{\text{Ga}}$-$\text{O}_{\text{N}}$ in InGaN alloy.} (a) Formation energy of $\text{V}_{\text{Ga}}$-$\text{O}_{\text{N}}$ in InGaN as functions of Fermi level reported in Ref. \cite{Alkauskas2016PRB}. (b) Calculated $C_{\text{tot}}$ and $A$ coefficient as a function of excess carrier density in intrinsic InGaN. The blue line shows the calculated $A$ coefficient when carrier emission is included, the red line shows the calculated $A$ coefficient when carrier emission is omitted, and the green line shows the calculated $A$ coefficient when recombination at shallow level is neglected. (c, d) Calculated $A$ coefficient changing with equilibrium Fermi level and excess carrier density when carrier emission is included and omitted, respectively.}\label{Fig4}
\end{figure}

\par The defect configuration studied in Ref. \cite{Alkauskas2016PRB} is a vacancy-substitution complex ($\text{V}_{\text{Ga}}$-$\text{O}_{\text{N}}$) in InGaN alloy ($E_g$ = 2.42 eV), which has three charge-state transition levels, $(0/+)$, $(0/-)$, and $(-/2-)$ levels, as plotted in Fig. \ref{Fig4}(a). We reproduce capture coefficients (Table S1) using Nonrad code \cite{Turiansky2021CPC} based on the parameters given in Ref. \cite{Alkauskas2016PRB} to be fully consistent with their calculation, and show the related configuration coordinate diagrams in Fig. S2 and Fig. S3 of Supplemental Material \cite{SM}. If the $C_{\text{tot}}$ formula is simply adopted, the corresponding $A$ coefficient is only $10^{-8}$ s$^{-1}$ (assuming defect density as $10^{16}$ cm$^{-3}$), which indicates $\text{V}_{\text{Ga}}$-$\text{O}_{\text{N}}$ is not an effective recombination defect in InGaN, consistent with the conclusion of Ref. \cite{Alkauskas2016PRB} that the complex in the ground state cannot act as a recombination center. However, it should be noted that both the $(0/-)$ and $(-/2-)$ levels are very close to the CBM, so the carrier emission from these two levels cannot be neglected according to our previous analysis.

\par Fig. \ref{Fig4}(b) shows the calculated $A_{\text{EIR}}$ of InGaN when $n_0=p_0$ (equilibrium Fermi level is at the middle of the band gap). When the excess carrier density is not very high, the electron emission rate $C_n^0 n_2$ (emission from $(0/-)$ level) and $C_n^- n_1$ (emission from $(-/2-)$ level) are both non-negligible. The strong carrier emission increases the defect density in neutral state in a transition path $-2\rightarrow -1 \rightarrow 0$, in stark contrast to the claim in Ref. \cite{Alkauskas2016PRB} that almost all the defects stay in $-1$ and $-2$ charge states, as shown in the comparison in Fig. S4. With the high density of neutral defects, the quick electron-hole recombination through the deep $(0/+)$ level makes the $A_{\text{EIR}}$ coefficient as high as $10^5$ s$^{-1}$ at low $\Delta n$. However, the value of $A_{\text{EOR}}$ is lower than $10^{-7}$ s$^{-1}$, indicating a huge error caused by omitting the carrier emission. It is also reported in Ref. \cite{Alkauskas2016PRB} that the slow hole capture process at shallow $(-/2-)$ level acts as a “rate-limiting” step that retards the overall nonradiative recombination. Our results here show clearly that its large electron emission rate prevents the overall recombination rate from being limited by the “rate-limiting” step, because the quick electron emission makes most defects transit back to the neutral state.

\par When $\Delta n$ increases, the competition between emission rate and capture rate at both $(0/-)$ and $(-/2-)$ levels changes the distribution of the defect densities in different charge states, that is, $\text{V}_{\text{Ga}}$-$\text{O}_{\text{N}}$ transits from $-1$ and $0$ charge states to $-2$ charge state gradually. Then the density decrease of neutral defects and the density increase of $-1$ and $-2$ charged defects cause the decrease of $A$ coefficient as $\Delta n$ increases, as shown by the blue line in Fig. \ref{Fig2}(b), because the role of slow recombination through the shallow $(0/-)$ and $(-/2-)$ levels becomes increasingly important. Comparing the declining blue line (emission included) and the constant red line (emission omitted) line, we can see that the error of the calculated $A$ coefficient is reduced from 13 orders of magnitude to 0 as $\Delta n$ increases. Under extremely high $\Delta n$ over $10^{22}$ cm$^{-3}$, the electron capture at $(-/2-)$ level prevails over the electron emission overwhelmingly, resulting in a much higher density of $-2$ charged defects. Then the slow hole capture process of shallow $(-/2-)$ level becomes the rate-limiting step, similar to the results calculated without emission \cite{Alkauskas2016PRB}, so the error in $A_{\text{EOR}}$ also disappears.

\par Fig. \ref{Fig4}(c, d) show the heat map of $A$ coefficients as functions of equilibrium Fermi level (as the Fermi level shifts from left to right, the conductivity changes from $p$-type to intrinsic and $n$-type) and excess carrier densities in InGaN. When Fermi level ranges from about 0.6 eV above the VBM to the CBM, the calculated $A_{\text{EOR}}$ remains unchanged against the increase of $\Delta n$ and is as small as $10^{-8}$ s$^{-1}$, but $A_{\text{EIR}}$ within this Fermi level range undergoes a gradual decrease with the increase of $\Delta n$. In strong $p$-type condition with Fermi level below 0.5 eV, both $A_{\text{EOR}}$ and $A_{\text{EIR}}$ are large when $\Delta n < 10^{16}$ cm$^{-3}$, because most defects stay in $+1$ charge state in both methods and thus the rapid recombination at $(0/+)$ level facilitates the overall recombination. Nevertheless, the $A$ coefficient causes non-negligible errors if the carrier emission is omitted regardless of the equilibrium Fermi level position.

\par In Ref. \cite{Alkauskas2016PRB}, the omission of carrier emission causes the small $A$ coefficient and slow recombination rate, so that an efficient recombination cycle would not be accomplished if only the ground state of $\text{V}_{\text{Ga}}$-$\text{O}_{\text{N}}$ is considered, and the authors resorted to the assistance of excited state for explaining the fast recombination. After revisiting the process with carrier emission considered, we find the $A$ coefficient of the ground-state $\text{V}_{\text{Ga}}$-$\text{O}_{\text{N}}$ can also be large, so its role in recombination should not be neglected. Similar effects of carrier emission may also exist in the recombination assisted by excited states.

\begin{figure}[htbp]
\centering
\includegraphics[width=0.5\textwidth]{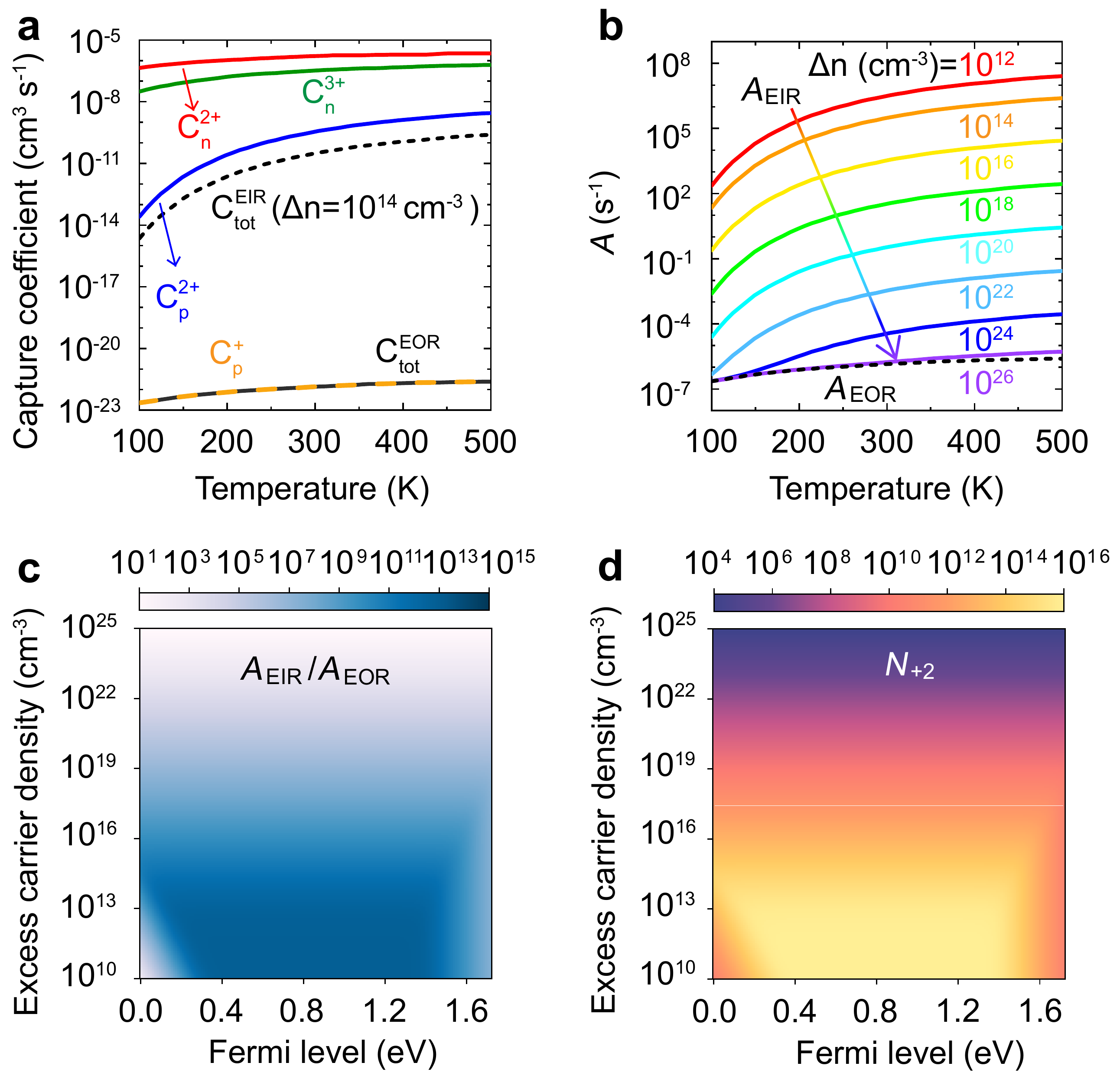}
\caption{\textbf{Effects of carrier emission on the nonradiative recombination assisted by $\text{Pb}_\text{I}$ defect in CsPbI$_3$.} (a) Temperature-dependent capture coefficients at $(+1/+2)$ and $(+2/+3)$ levels (extract from Ref. \cite{Zhang2020JPCC}) and the total capture coefficients $C_{\text{tot}}$ with carrier emission effect omitted and with carrier emission effect included. (b) Temperature-dependent $A_{\text{EIR}}$ (solid) coefficients with varied excess carrier density and $A_{\text{EIR}}$ (dashed) coefficients. (c) Heat map of $A_{\text{EIR}}$/$A_{\text{EOR}}$ ratio as functions of excess carrier density and Fermi level. (d) Heat map of defect density of $\text{Pb}_\text{I}$ in +2 charge state of as functions of excess carrier density and Fermi level, when carrier emission is included.}\label{Fig5}
\end{figure}

\par In another example, carrier recombination at $\text{Pb}_\text{I}$ in halide perovskite CsPbI$_3$, our study shows omitting the carrier emission also underestimates the $A$ coefficient by orders of magnitude. In Ref. \cite{Zhang2020JPCC}, the antisite $\text{Pb}_\text{I}$ was shown to have $(+1/+2)$ and $(+2/+3)$ transition levels. Using the $C_{\text{tot}}$ formula, it has been shown in Ref. \cite{Zhang2020JPCC} that it is the slow hole capture at shallow (+1/+2) level with a capture coefficient $C_p^+$ smaller than $10^{-21}$ cm$^{3}$ s$^{-1}$ that limits the overall recombination rate. As a result, the emission-omitted total capture coefficient $C_{\text{tot}}^{\text{EOR}}$ was found to be very small, arriving at the conclusion that $\text{Pb}_\text{I}$ cannot act as a recombination center. However, the $(+1/+2)$ level lies just 0.33 eV below CBM, so the carrier emission from the level should play a role in the nonradiative recombination. To make a comparison of capture coefficients, an emission-included total capture coefficient can be defined,
\begin{equation}\label{Ctot_EIR}
    C_{\text{tot}}^{\text{EIR}}=\frac{A_{\text{EIR}}}{N_{\text{tot}}}.
\end{equation}

\par In Fig. \ref{Fig5}(a), we compare our calculated $C_{\text{tot}}^{\text{EIR}}$ (at $\Delta n=10^{14}$ cm$^{-3}$) and $C_{\text{tot}}^{\text{EOR}}$ (independent of $\Delta n$) calculated in Ref. \cite{Zhang2020JPCC}. As we see, the difference can be as large as 10 orders of magnitude. If the defect density is assumed to be $10^{16}$ cm$^{-3}$, we calculate $A_{\text{EIR}}$ with different excess carrier densities $\Delta n$ and $A_{\text{EOR}}$ in Fig. \ref{Fig5}(b). When $\Delta n$ reaches an extremely high (but not reasonable) value of $10^{26}$ cm$^{-3}$, $A_{\text{EIR}}$ decreases to a lower limit and becomes comparable to $A_{\text{EOR}}$, and the corresponding $C_{\text{tot}}^{\text{EIR}}$ coincides with $C_{\text{tot}}^{\text{EOR}}$ of $10^{-22}$ cm$^{3}$ s$^{-1}$ reported in Ref. \cite{Zhang2020JPCC}.

\par In Fig. \ref{Fig5}(c), we plot the ratio of $A_{\text{EIR}}$/$A_{\text{EOR}}$ (equals $C_{\text{tot}}^{\text{EIR}}$/$C_{\text{tot}}^{\text{EOR}}$) at different excess carrier densities and Fermi levels. The large values of the ratio show $A_{\text{EIR}}$ are much larger than $A_{\text{EOR}}$ in the whole range of Fermi level. The difference between $A_{\text{EIR}}$ and $A_{\text{EOR}}$ is very large when $\Delta n$ is low and becomes relatively smaller as $\Delta n$ increases. However, even at an extremely high ∆n of $10^{22}$ cm$^{-3}$, the difference between $A_{\text{EOR}}$ and $A_{\text{EIR}}$ can still be as large as 3 orders of magnitude. Since $A_{\text{EOR}}$ does not change with $\Delta n$, the change of $A_{\text{EIR}}$/$A_{\text{EOR}}$ ratio originates mainly from the variation of $A_{\text{EIR}}$ with $\Delta n$, which is determined by the densities of $\text{Pb}_\text{I}$ in +2 charge state. As shown in Fig. \ref{Fig5}(d), when carrier emission is considered, the defect density in +2 charge state significantly increases when $\Delta n$ decreases, so $A_{\text{EIR}}$ is relatively large at low $\Delta n$ but gradually decreases with the increase of $\Delta n$. According to our calculation, when $\Delta n$ is on the order of $10^{14}$ cm$^{-3}$ at 300 K, $A_{\text{EIR}}$ is about $10^{5}$ s$^{-1}$ when defect density is assumed to be $10^{16}$ cm$^{-3}$. This value increases to $10^{7}$ s$^{-1}$ if we a assume a higher defect density of $10^{18}$ cm$^{-3}$, indicating that $\text{Pb}_\text{I}$ may act as an effective recombination center in CsPbI$_3$. However, if the carrier emission is omitted in the calculation, $A_{\text{EOR}}$ is only about $10^{-6}$ s$^{-1}$ and $10^{-4}$ s$^{-1}$ at defect density of $10^{16}$ cm$^{-3}$ and $10^{18}$ cm$^{-3}$ respectively, indicating that $\text{Pb}_\text{I}$ can be excluded as a recombination center. Therefore, omitting carrier emission causes an incorrect conclusion about the role of $\text{Pb}_\text{I}$ in nonradiative recombination.

\par Apart from $\text{V}_{\text{Ga}}$-$\text{O}_{\text{N}}$ in InGaN and $\text{Pb}_\text{I}$ in CsPbI$_3$, we also examined another multi-level defect and got similar conclusions in Fig. S5 of Supplemental Material \cite{SM}, where we compared our calculated $A_{\text{EIR}}$ and $A_{\text{EOR}}$ based on the original data from the recent study \cite{Moustafa2024PRM}.

\subsection{\label{sec:level2}Parallel-circuit recombination model of multi-level defects}

\par
\begin{figure*}[htbp]
\centering
\includegraphics[width=0.8\textwidth]{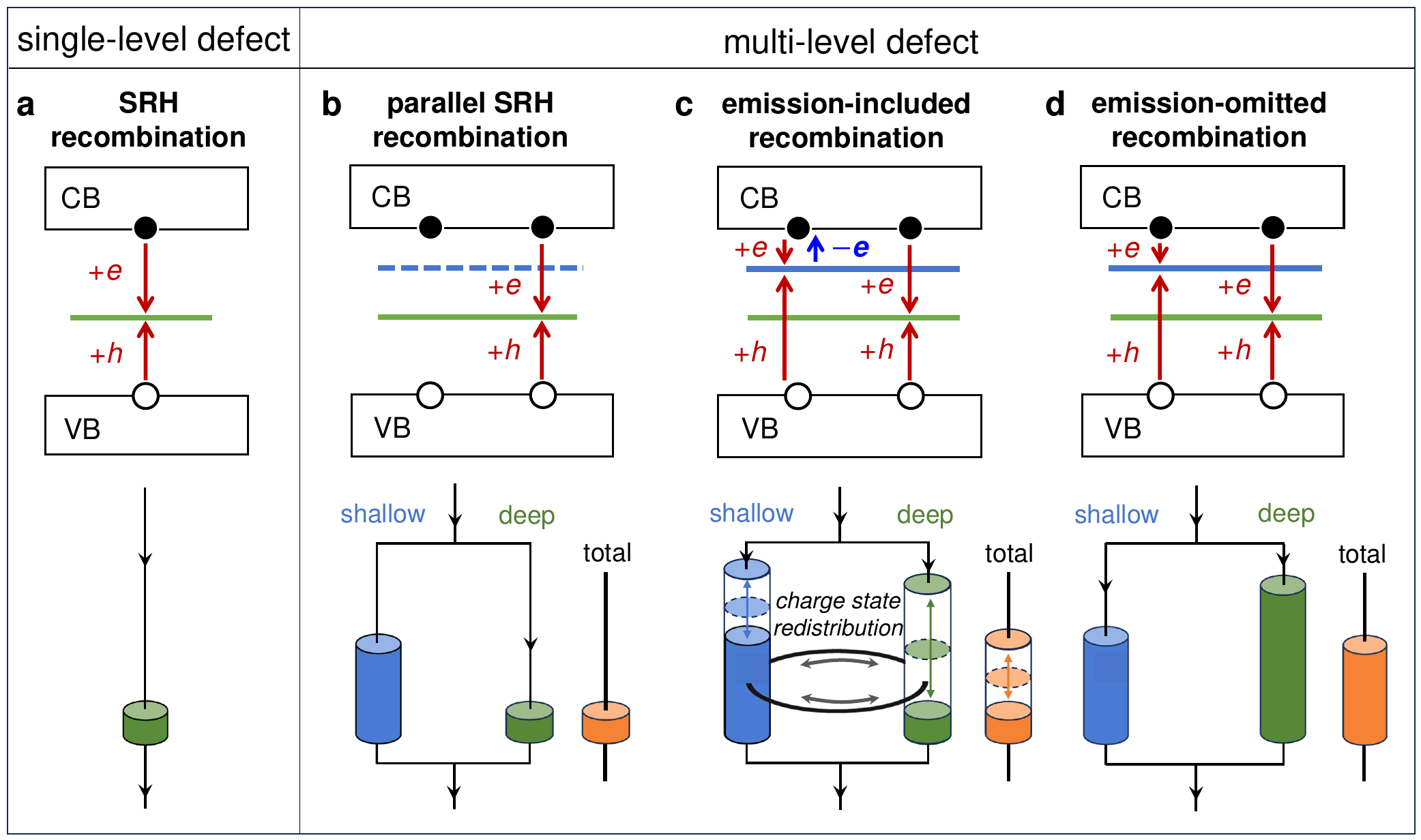}
\caption{\textbf{Schematic plot of single-resistor circuit model of single-level defect assisted carrier recombination and two-resistor parallel-circuit models of two-level defect assisted recombination.} The recombination rates are analogized to electrical currents in parallel circuits, and the electrical resistances (represented by the length of the resistors) are analogized to the carrier lifetime $\tau$ ($\tau=1/A$). The longer the resistance looks, the longer $\tau$ is and the smaller the $A$ coefficient is. For the recombination assisted by a single-level defect as shown in (a), the carrier lifetime $\tau$ is represented by a resistor of the circuit. For the recombination assisted by a two-level defect in (b-d), it is analogized to a parallel circuit with two resistors. The SRH recombination lifetime $\tau_1$ through the shallow level is represented by the resistor in blue, the lifetime $\tau_2$ through the deep level is represented by the resistor in green, and the total lifetime $\tau_{\text{tot}}$ is represented by the resistor in orange.}\label{Fig6}
\end{figure*}

 For a semiconductor with multiple types of single-level defects, the total recombination rate is sum of the SRH recombination rates of all types of defects. Correspondingly, the total carrier lifetime $\tau _{\text{tot}}$ can be written as,
\begin{equation}\label{lifetime}
    \frac{1}{\tau _{\text{tot}}}=\frac{1}{\tau_1}+\frac{1}{\tau_2}+\frac{1}{\tau_3}+\cdots,
\end{equation}
where $\tau_1$, $\tau_2$ and $\tau_3$ are the carrier lifetime in the semiconductor with only one type of single-level defects. This relation can be analogized to the total resistance of multiple resistors in parallel, where the lifetime τ is analogized to the resistance and the recombination coefficient $A$ is analogized to the electric current (Note $\tau=1/A$ for the SRH recombination assisted by a single-level defect). The resistor analogue of a single-level defect is schematically shown in Fig. \ref{Fig6}(a). The two-resistor parallel-circuit analogue of two single-level defects is schematically shown in Fig. \ref{Fig6}(b).

\par For multi-level defects, if we assume each defect level assists the SRH recombination of conduction-band electrons and valence-band holes independently and do not influence each other, then the multi-level defect can be approximated as equivalent to multiple independent single-level defects. We call this approximation as the parallel SRH (PSRH) recombination model of multi-level defects. For example, we can approximate the two-level defect in Fig. \ref{Fig2}(a) as equivalent to two single-level defects, as plotted in Fig. \ref{Fig6}(b). The first single-level defect with the shallow $(0/-1)$ level can be taken as a long resistor because the recombination coefficient $A_1$ through the level is low and the corresponding lifetime $\tau_1$ is long, and the second single-level defect with the deep $(0/+1)$ level can be taken as a short resistor because the recombination coefficient $A_2$ through the level is high and the corresponding lifetime $\tau_2$ is short. Then, similar to the total resistance of resistors in parallel, the total lifetime $\tau_{\text{tot}}$ is short as determined mainly by the short $\tau_2$. In other words, since $\tau_1$ is much longer than $\tau_2$, the $\tau_1$ influence of the shallow level can be neglected. Based on this approximation, several recent studies \cite{Decock2010JAP,Ricca2021PRR,Vedel2023PRB} on multi-level defects considered only the deep level of the defect and neglected the shallow level when calculating the carrier recombination coefficient and lifetime. 

\par Approximating a multi-level defect to multiple independent single-level defects is also commonly adopted in semiconductor device modeling software, such as the well-known Sentaurus and Silvaco TCAD \cite{Xiang2024IEEEJP}. However, as we mentioned, the $ABC$ recombination model with a constant $A$ coefficient is valid for single-level defects, so the PSRH approximation, which takes the sum of $A$ coefficients of single-level defects as $A$ of a multi-level defect, will cause an incorrect conclusion that $A$ of a multi-level defect is also a constant independent of excess carrier density, in contrast with the $A(n)BC$ model. Now we will analyze why the PSRH approximation causes errors in the calculated recombination coefficient and thus causes the incorrect conclusion.

\par Using the PSRH approximation, we calculated $A_{\text{PSRH}}$ and the corresponding $\tau_{\text{tot}}$ for $\text{V}_{\text{Ga}}$-$\text{O}_{\text{N}}$ in InGaN, and found that $A_{\text{PSRH}}$ and $\tau_{\text{tot}}$ do not change with $\Delta n$, as shown by the flat line in Fig. \ref{Fig4}(b). In a wide range of $\Delta n$, $A_{\text{PSRH}}$ is much lower higher than $A_{\text{EIR}}$. Therefore, neglecting the influence of shallow level in the PSRH approximation causes large errors in the calculated recombination coefficient $A$ and lifetime $\tau_{\text{tot}}$. Comparing $A_{\text{PSRH}}$, $A_{\text{EIR}}$ and $A_{\text{EOR}}$ in Fig. \ref{Fig4}(b), we can find that $A_{\text{EIR}}$ decreases from $A_{\text{PSRH}}$ to $A_{\text{EOR}}$ as $\Delta n$ increases, i.e., $A_{\text{PSRH}}$ is the upper limit of $A_{\text{EIR}}$ at low $\Delta n$, while $A_{\text{EOR}}$ is the lower limit of $A_{\text{EIR}}$ at high $\Delta n$. That indicates the origin of the error in the calculated $A_{\text{PSRH}}$ should be different from that in $A_{\text{EOR}}$.

\par To understand the origin, we propose a modified parallel-circuit recombination model that includes the carrier emission effect (Fig. \ref{Fig6}(c)) and a model that omits the carrier emission effect (Fig. \ref{Fig6}(d)). Compared to the PSRH approximation in Fig. \ref{Fig6}(b) that takes a multi-level defect as equivalent to multiple single-level defects, the major modification in the emission-included parallel-circuit recombination model is that the lifetimes $\tau_1$ and $\tau_2$ of the two assumed single-level defects are not constants, and they are correlated with each other and can change with $\Delta n$. In contrast, $\tau_1$ and $\tau_2$ of two single-level defects in the PSRH approximation are constants and not influenced by other each other, which further makes the total lifetime ($\tau_{\text{tot}} \approx \tau_2$) a constant. The necessity of the modification results from the defect charge state redistribution with varied excess carrier density $\Delta n$ when the carrier emission effect is included. As proved in Supplemental Material, the total carrier recombination rate assisted by multi-level defect can be expressed as the sum of recombination rate through each defect level, so the total lifetime can indeed be described by Eq. (\ref{lifetime}). However, the recombination coefficients and lifetimes of each defect level should depend strongly on the defect densities in each charge state, which varies sensitively with $\Delta n$ due to the competition between carrier capture and carrier emission, as analyzed in Section 2. For example, $A_1$ and $\tau_1$ of the shallow level and $A_2$ and $\tau_2$ of the deep level in Fig. \ref{Fig6}(c) should all vary with $\Delta n$, which further makes the total lifetime $\tau_{\text{tot}}$ vary with $\Delta n$, so the length of two parallel resistors and the length of total resistor are all variable. Since the total density of the defect in all three charge states is fixed, the recombination coefficients ($A_1$ and $A_2$) and lifetimes ($\tau_1$ and $\tau_2$) of two defect levels are correlated with each other, rather than being independent.

\par In the PSRH approximation, since the SRH recombination through the two assumed single-level defects are independent of each other and the density of each single-level defect is assumed the same as the density of the original two-level defects, the $\Delta n$ and defect-charge-state-distribution mediated correlation between $\tau_1$ and $\tau_2$ (similarly for $A_1$ and $A_2$) disappears. Since the density of the single-level defect with the deep $(0/+1)$ level is fixed at a high value, all the defects are assumed to take the $0$ and $+1$ charge states, which will cause rapid recombination through the deep $(0/+1)$ level and give a large constant $A_2$ and thus a small $\tau_2$. Therefore, the total lifetime ($\tau_{\text{tot}} \approx \tau_2$) calculated using PSRH approximation is also a constant and underestimated (as shown by the short resistor in Fig. \ref{Fig6}(b)) seriously by the incorrect defect density assumption. According to this, we can also understand why $A_{\text{PSRH}}$ is the upper limit of $A_{\text{EIR}}$ at low $\Delta n$ in Fig. \ref{Fig4}(b), because most of the defects indeed take the $0$ and $+1$ charge states at low $\Delta n$. As $\Delta n$ increases, more defects transit into the $-1$ state, so the error caused by PSRH approximation increases. Such errors should be paid special attention to when one or several shallow levels of a multi-level defect is neglected in the carrier dynamics calculation. 

\par For the emission-omitted parallel-circuit recombination model in Fig. \ref{Fig6}(d), $\tau_1$ and $\tau_2$ (similarly for $A_1$ and $A_2$) are also constants independent of $\Delta n$ and not correlated with each other. Omitting carrier emission means no competition between carrier capture and carrier emission, so the defect densities in different charge states are constants independent of $\Delta n$ as shown in Fig. \ref{Fig2}(f), which makes $\tau_1$ and $\tau_2$ constants independent of $\Delta n$. Meanwhile, since omitting carrier emission make most of the defects stuck in the $-1$ state and the densities of $0$ and $+1$ charged defects are very low, $\tau_2$ of the $(0/+1)$ level becomes much longer. Then both $\tau_1$ and $\tau_2$ are long, so $\tau_{\text{tot}}$ is also long and the corresponding $A_{\text{EOR}}$ is small, explaining why $A_{\text{EOR}}$ is the lower limit of $A_{\text{EIR}}$ at high $\Delta n$ in Fig. \ref{Fig4}(b).

\section{\label{sec:level1}Conclusions}

\par In summary, we find the carrier emission effect plays important roles in multi-level defect assisted recombination and cannot be neglected in the calculation of recombination rates. Neglecting the effect, as in recent calculation studies based on the $C_{\text{tot}}$ formula, can cause the independence of the recombination coefficient $A$ on the excess carrier density and serious deviation of $A$, especially when one of the defect levels is shallow, because the defect will be stuck in a charge state after the shallow level capture carriers quickly and cannot assist electron-hole recombination. When the effect is considered, the competition between carrier capture and carrier emission makes the distribution of defect densities in different charge states depend on the excess carrier density, thus making the $A$ coefficient become a function of the carrier density $n$. Therefore, the well-known $ABC$ recombination rate model which takes the defect assisted SRH recombination coefficient $A$ as a constant should be reformed into the $A(n)BC$ model in which the multi-level defect assisted recombination coefficient $A(n)$ is a function of excess carrier density.

\par In many carrier-dynamics experiments such as those using time-resolved photoluminescence, the $ABC$ recombination rate model is adopted to fit the excess carrier density decay and $A$ is taken as a constant describing the defect contribution to the recombination. Our current work indicates that the fitting may be inaccurate and the influences of defects may be misleading when the recombination is assisted by multi-level defects. Furthermore, when using the $ABC$ model to calculate the quantum efficiency in the design of optoelectronic devices such as solar cells and LEDs, the findings in this work should also be considered for the accurate calculation of the $A(n)$ function and the optimization of the device performance with varied light illumination intensity, carrier injection and thus different excess carrier densities.

\section{\label{sec:level1}Data availability}

The data supporting the key findings of this article are available within the article and the Supplemental Material file. 

\section{\label{sec:level1}Code availability}

The codes used to calculate the recombination coefficients are available from the corresponding author upon reasonable request.

\begin{acknowledgments}
S. Wang and M. Huang contributed equally to this work. This work was supported by National Natural Science Foundation of China (12334005, 12188101, 12174060 and 12404089), National Key Research and Development Program of China (2022YFA1402904 and 2024YFB4205002), Science and Technology Commission of Shanghai Municipality (24JD1400600), China Postdoctoral Science Foundation Project (2023M740722), China National Postdoctoral Program for Innovative Talents (BX20230077) and Project of MOE Innovation Platform.
\end{acknowledgments}

\bibliography{apssamp}

\end{document}